\documentclass[12pt]{iopart}
\usepackage{iopams}  

\usepackage{bm}
\usepackage{cite}
\usepackage{graphicx}

\newcommand{\bra}{\left\langle}
\newcommand{\ket}{\right\rangle}
\newcommand{\pder}[2]{\frac{\partial #1}{\partial  #2}}

\newcommand{\der}[2]{\frac{\rmd #1}{\rmd  #2}}
\newcommand{\tder}[2]{\frac{\rmd^2 #1}{\rmd  #2^2}}
\newcommand{\bv}[1]{{\boldsymbol #1}}

\newcommand{\rhoi}{\rho_{\rm i}}
\newcommand{\rhof}{\rho_{\rm f}}
\newcommand{\nint}{n_{\rm i}}
\newcommand{\nf}{n_{\rm f}}

\begin{document}

%\hfill {\bf ver. 091209}\\
\hfill KEK-TH-1343 

\title{Two Langevin equations in the Doi-Peliti formalism}

\author{Kazunori Itakura}
\address{
KEK Theory Center, IPNS, 
High Energy Accerelator Research Organization,
1-1 Oho, Tsukuba, Ibaraki, 305-0801, Japan, and \\
Department of Particle and Nuclear Studies, 
%School of High Energy Accelerator Science, 
Graduate University for Advanced Studies (SOKENDAI), 
1-1 Oho, Tsukuba, Ibaraki, 305-0801, Japan
}
\ead{kazunori.itakura@kek.jp}

\author{Jun Ohkubo}
\address{
Institute for Solid State Physics, University of Tokyo, 
Kashiwanoha 5-1-5, Kashiwa, Chiba 277-8581, Japan
}
\ead{ohkubo@issp.u-tokyo.ac.jp}

\author{Shin-ichi Sasa}
\address{
Department of Pure and Applied Sciences,
University of Tokyo, Komaba, Tokyo 153-8902, Japan
}
\ead{sasa@jiro.c.u-tokyo.ac.jp}

\begin{abstract}
A system-size expansion method is incorporated into the Doi-Peliti 
formalism for stochastic chemical kinetics. The basic idea of the 
incorporation is to introduce a new decomposition of unity 
associated with a so-called Cole-Hopf transformation. This approach  
elucidates a  relationship between two different Langevin 
equations; one is associated with a coherent-state path-integral 
expression and the other describes density fluctuations.
A simple reaction scheme  $X \rightleftarrows X+X$ 
is investigated as an illustrative example. 
\end{abstract}

\vspace{-5mm}

%Uncomment for PACS numbers title message
\pacs{05.10.Gg,05.40.-a,82.20.-w}
% stochastic analysis method
% fluctuation phenomena
% chemical kinetics and dynamics

% Keywords required only for MST, PB, PMB, PM, JOA, JOB? 
%\vspace{2pc}
%\noindent{\it Keywords}: Article preparation, IOP journals
% Uncomment for Submitted to journal title message
%\submitto{\JPA}
% Comment out if separate title page not required
%\maketitle

%%%%%%%%%%%%%%%%%%%%%%%%%%%%%%%%%%%%%%%%%
\section{Introduction}                  %
%%%%%%%%%%%%%%%%%%%%%%%%%%%%%%%%%%%%%%%%%

% Size expansion

System-size expansion methods were developed in the analysis
of master equations for stochastic chemical kinetics 
\cite{Kampen,KMK,Gardiner_book}. 
In these methods, under the assumption of an extensive property 
of macroscopic quantities, fluctuation effects 
are systematically taken into account by assuming the inverse 
of a dimensionless system size to be a small perturbation parameter.
The leading-order description of a system is provided as 
a deterministic equation for a density variable 
due to the law of large numbers, and macroscopic 
fluctuations are described by a  Langevin equation
which reduces to the deterministic equation in the limit
of an infinite system size. 

% Doi-Peliti 

The Doi-Peliti formalism is known as another standard 
technique, where master equations are equivalently expressed 
in terms of a set of creation and annihilation operators 
\cite{Doi1976,Doi1976a}. 
Since the original proposal, many models have been 
analyzed within this formalism, as seen in a list of papers quoted in 
Ref. \cite{Mattis1998}. 
In particular, a coherent-state path-integral expression
is useful for theoretical study, because it takes a form 
similar to that for a continuum field theory 
\cite{Peliti1985,Mattis1998,Tauber2005}. 

% question 

Then, in some cases including an example discussed in
this paper, the dynamical  action of a coherent-state
path-integral expression becomes equivalent to that of 
a Langevin equation. However, since this Langevin equation 
does not coincide with that obtained by system-size 
expansion methods, it does not directly describe the stochastic 
time evolution of a density. Note that this difference is 
not recognized well; indeed, as presented in a review 
paper \cite{Panja2004}, a Langevin equation obtained 
by a coherent-state path-integral expression \cite{Pechenik1999} 
has been analyzed in the problems of fluctuating front 
propagation of some reaction-diffusion systems 
as if it describes the time evolution of a density. 
(See however a  different 
argument in Ref. \cite{Doering2003}.)

% presentation of the questions and what we did

In order to resolve this confusing situation, in the 
present paper, we develop a system-size expansion method 
for a coherent-state path-integral expression in the 
Doi-Peliti formalism.  Our consideration consists 
of three steps. In the first place, we express
statistical quantities in terms of a variable that is 
straightforwardly introduced in the coherent-state 
path-integral expression. In addition to the presentation 
of such an expression, which was already discussed in 
Refs. \cite{Pechenik1999,Tauber2005}, we shall provide 
a clear argument on the indirect correspondence between 
density fluctuations and trajectories of the 
Langevin equation associated with the coherent-state 
path-integral expression. At the second step, we seek 
for a simple expression
of density fluctuations in terms of coherent states.
Although it has been argued that such an expression
would be realized by a so-called Cole-Hopf transformation 
\cite{Janssen2001,Frederic2001,Andreanov2006,Lefevre2007}, 
the description involves non-transparent procedures, 
as will be indicated in the beginning of section \ref{sec:new}.
Our main idea in the present paper is to introduce 
a new decomposition of unity associated with the 
Cole-Hopf transformation, by which we obtain a new 
path-integral expression. Finally, this expression
enables us to perform a system-size expansion,
and it directly leads to the Langevin equation 
describing density fluctuations.

% road map

This paper is organized as follows. 
In section \ref{sec:model}, we introduce a model we study 
within the Doi-Peliti formalism. In order to make our presentation
as instructive and transparent as possible,
we focus on a simple reaction scheme 
$ X \rightleftarrows X+X$ in a single box under 
an assumption
that diffusion processes are sufficiently fast.
In section \ref{sec:coherent_path}, we review a 
coherent-state path-integral expression, and derive 
the first type Langevin equation. 
In section \ref{sec:new}, we  propose a new decomposition 
of unity, by which we derive the second type Langevin 
equation. In section \ref{sec:diff}, we briefly
discuss effects of diffusion.  

%%%%%%%%%%%%%%%%%%%%%%%%%%%%%%%%%%%%%%%%%%%%%%%%%
\section{Model}\label{sec:model}                %
%%%%%%%%%%%%%%%%%%%%%%%%%%%%%%%%%%%%%%%%%%%%%%%%%

% situation and master equation

Let $\Omega$ be a volume of a box in which particles 
are confined. We consider a chemical reaction 
$ X+ X \rightleftarrows X$; the duplication reaction 
$X \to X+X$ occurs with ratio $\alpha$ for each particle, 
and its backward reaction $X+X \to X$ occurs with 
ratio $\beta$ when two particles are close to
each other less than a reaction length $r_0$.
We assume that a diffusion time %over the system size 
is much smaller than typical time scales of the 
reaction: $\alpha^{-1}$ and $\beta^{-1}$,
and ignore the effect of diffusion for the time 
being (we will briefly discuss it in section \ref{sec:diff}). 
Then, since the probability of finding another particle 
within a reaction region for a given particle is 
$ 4\pi r_0^3/(3\Omega)$, the rate of the reaction 
$X+X \to X$ for each pair of particles becomes
$\beta 4\pi r_0^3/(3\Omega)$. In the argument below, 
a dimensionless quantity $ 4\pi r_0^3/(3\Omega)$
is replaced with $1/\Omega$, for  notational simplicity. 
Now, let $P_n(t)$ be a probability of finding $n$ particles 
at time $t$. The time evolution of $P_n(t)$ is 
described by the master equation
\begin{eqnarray}
\fl\quad
\frac{\partial}{\partial t} P_n(t) = \alpha [(n-1)P_{n-1}(t) - n P_n(t)]
+ \frac{\beta}{\Omega} [n (n+1) P_{n+1}(t) - n(n-1) P_n(t)]\, ,
\label{master_eq}
\end{eqnarray}
where $P_{-1}(t)\equiv 0$.

% time evolution in the operator formalism

State of the system is specified by a vector $|n \rangle $,
which represents a situation that there are  $n$ particles in 
the system. Let $V$ be a vector space 
spanned by an orthogonal set: 
$V=\{\, \vert n \rangle\, \vert\, n=0,\cdots, \infty \}$. 
A remarkable idea of the
Doi-Peliti formalism \cite{Doi1976,Doi1976a} is that 
a series of infinite number of $P_n(t)$ is collectively 
treated as a single vector in $V$:
\begin{eqnarray}
| \psi(t) \rangle 
= \sum_{n=0}^\infty P_n(t) | n \rangle\, .
\end{eqnarray}
Then, the master equation (\ref{master_eq}) is equivalently 
rewritten in a compact form
\begin{eqnarray}
\frac{\partial}{\partial t} | \psi(t) \rangle 
= -\hat{H} | \psi(t) \rangle,
\label{Schrodinger_like_eq}
\end{eqnarray}
where $\hat{H}$ is an infinite-dimensional matrix.
This matrix is expressed in terms of two matrices 
$\hat{a}$ and $\hat{a}^\dagger$  defined by 
$ \hat{a} | n +1\rangle = (n+1) | n  \rangle$ 
and $ \hat{a}^\dagger | n  \rangle = | n + 1 \rangle$
for any $n$  (with $ \hat{a} | 0 \rangle = 0$): 
\begin{eqnarray}
\hat{H} =
- \alpha (\hat{a}^\dagger - 1) \hat{a}^\dagger \hat{a} 
- \frac{\beta}{\Omega} (1 - \hat{a}^\dagger) 
\hat{a}^\dagger \hat{a}^2\, .
\label{time_evolution_operator}
\end{eqnarray}
The two matrices satisfy the bosonic commutation relations
$[\hat{a}, \hat{a}^\dagger] = 1$ and 
$ [\hat{a},\hat{a}] = [\hat{a}^\dagger,\hat{a}^\dagger] = 0$.
Since $\hat a^\dag$ and $a$ are similar to the 
creation and annihilation operators, and $\hat H$ is an analog 
of the Hamiltonian, the Doi-Peliti formalism is also called the 
second-quantization method.

%% bra-space  

For any $|v_1 \rangle $ and $|v_2 \rangle $ in $V$, 
the inner product $\bra v_1|v_2 \ket $ is naturally 
defined by noting $\langle m | n \rangle = n!\, \delta_{m,n}$.
There is a special vector $|\mathcal{P} \rangle$ such 
that $\bra  n |\mathcal{P} \ket=1$ for any $n$, 
which is employed to express the expectation value of 
observables. This vector is explicitly written as 
$|\mathcal{P} \rangle = \rme^{\hat{a}^\dagger}|0 \rangle$.
Let $A(x)$ be a polynomial of a variable $x$. 
We consider the  expectation value of  $A(n)$ at 
time $t=\tau$, which is denoted by $\bra A(n(\tau)) \ket$.
By using the %special 
vector $| \mathcal{P} \rangle $, 
we write
$\bra A(n(\tau)) \ket \equiv  
\sum_{n=0}^\infty A(n) P_n(\tau)
= \langle \mathcal{P} | {A}(\hat{a}^\dagger \hat{a}) | \psi(\tau) \rangle$.
In order to simplify this expression further, we define 
a polynomial $\mathcal{A}(x)$ associated with $A(x)$ by
the relation 
$\langle \mathcal{P} | A(\hat a^\dagger \hat{a})
=\langle \mathcal{P} | {\cal A}(\hat{a})$. 
That is, $\mathcal{A}(\hat a)$ is obtained by the {\it normal ordering}
of $A(\hat{a}^\dagger \hat{a})$.
As an example, we demonstrate for the simplest non-trivial 
case $A(x)=x^2$. Since 
$\langle \mathcal{P} | (\hat a^\dagger \hat a)^2
=\langle \mathcal{P} |  \hat a\hat a^\dagger \hat a
=\langle \mathcal{P} |  (\hat a^2+ \hat a)$, one finds  
$\mathcal{A}(x)=x^2+x$. More generally, 
by mathematical induction, one can prove that 
$\mathcal{A}(x)=x^k$ for $A(x)=x(x-1)\cdots (x-k+1)$ with
any integer $k$. 
The final expression of the expectation value of $A(n)$ at
$t=\tau$ becomes 
\begin{eqnarray}
\bra A(n(\tau)) \ket 
= \langle \mathcal{P} | {\cal A}(\hat{a}) | \psi(\tau) \rangle\, .
\label{eq_observables}
\end{eqnarray}

%%%%%%%%%%%%%%%%%%%%%%%%%%%%%%%%%%%%%%%%%%%%%%%%%%%%%%
\section{Coherent-state path-integral expression}    %
\label{sec:coherent_path}                            %
%%%%%%%%%%%%%%%%%%%%%%%%%%%%%%%%%%%%%%%%%%%%%%%%%%%%%%

% starting 

With the ``second-quantized'' expression of the master 
equation (\ref{Schrodinger_like_eq}), it is quite natural to 
consider the path-integral form of the expectation value 
(\ref{eq_observables}). 
In this section, we explain a procedure to obtain 
a path-integral expression by using coherent states 
%for (\ref{Schrodinger_like_eq}) 
within the model (\ref{time_evolution_operator}).
The result was already obtained in Ref. \cite{Pechenik1999}, 
but we shall provide a slightly different presentation 
of the derivation.  

% standard coherent state

Coherent states are eigenstates of the matrix $\hat a$. 
They are in general written as 
\begin{eqnarray}
| z  \rangle \equiv \rme^{z \hat{a}^\dagger} | 0 \rangle
\label{coherent:ket}
\end{eqnarray}
with a complex number parameter $z$ being the eigenvalue: 
$\hat a |z \rangle = z |z \rangle$. The corresponding
bra vector is denoted by $\langle z|$, that is, 
$\langle z|=\langle 0|\rme^{z^* \hat{a}}  $,
where $z^{*}$ is the complex conjugate of $z$.
The following decomposition of unity plays an  
essential role in constructing a path-integral 
expression with the coherent states:
\begin{eqnarray}
%\fl
\bm{1} = \sum_{n=0}^\infty \frac{1}{n!} | n \rangle \langle n |
= \sum_{n=0}^\infty \sum_{m=0}^\infty \frac{1}{n!} 
| n \rangle \langle m | \delta_{m,n} 
= 
\int \frac{\rmd^2 z}{\pi} \, \rme^{- |z|^2}
|z\rangle \langle z|\, ,
\label{identity_eq:standard}
\end{eqnarray}
where we have used 
\begin{eqnarray}
\delta_{n,m} =  \int \frac{\rmd^2 z}{\pi n!}\, 
\rme^{- |z|^2} z^{*m} z^n\, ,
\end{eqnarray}
with the integration measure 
$\rmd^2 z = \rmd(\mathrm{Re}\, z) \rmd(\mathrm{Im}\, z)$. 
See Refs. \cite{Mattis1998} and  \cite{Tauber2005} 
for the path-integral expression on the basis of 
(\ref{identity_eq:standard}).

% "different" decomposition of unity

Here, we notice a different decomposition of unity
\begin{eqnarray}
\bm{1} 
= \int_{-\infty}^\infty \rmd \phi 
\int_{-\infty}^\infty \frac{\rmd \varphi}{2\pi}
| \phi \rangle \langle -\rmi \varphi | \rme^{-\rmi \phi \varphi}\, ,
\label{identity_eq}
\end{eqnarray}
where we have used the formula
\begin{eqnarray}
\fl \qquad
\delta_{n,m} = \frac{1}{n!} \int_{-\infty}^\infty \rmd \phi \, 
\phi^n \left( - \frac{\rmd}{\rmd \phi}\right)^m \delta(\phi)
= \frac{1}{n!} \int_{-\infty}^\infty \rmd \phi \, 
\int_{-\infty}^\infty \frac{\rmd \varphi}{2\pi} 
\phi^n \left( \rmi \varphi\right)^m
\rme^{-\rmi \phi \varphi}\, .
\label{Kronecker_two_real}
\end{eqnarray}
Note that $\phi$ and $\varphi$ are {\it real} variables.
The path-integral expression on the basis of 
the decomposition of unity (\ref{identity_eq}) might be
less familiar, but we % wish to 
remind readers
that this expression was presented in the 
seminal paper by Peliti \cite{Peliti1985}.
It should be equivalent to the standard one obtained 
by using (\ref{identity_eq:standard}). Indeed, 
when we employ the expression with $(\phi, {\rm i}\varphi)$,
we can always move to the standard expression 
by formally replacing $(\phi, {\rm i}\varphi)$ with $(z,  z^*)$. 
In addition, as far as we understand, the expression
obtained by (\ref{identity_eq}) is more transparent
in the sense that the integration can be considered 
in an explicit manner. 
From these reasons, we review the coherent-state 
path-integral expression by using (\ref{identity_eq}).

% Coherent-state path-integral

The time-integration of \eref{Schrodinger_like_eq} yields 
\begin{eqnarray}
| \psi(\tau) \rangle 
= \lim_{\Delta t \to 0} 
[\exp(- \hat{H} \Delta t)]^{\tau/\Delta t}|\psi(0) \rangle\, .
\label{psit}
\end{eqnarray}
We assume that an initial condition obeys a Poisson distribution 
with its average $\bar{n}_0$. The initial state $| \psi(0) \rangle $
is then expressed as
\begin{eqnarray}
| \bar{n}_0; {\rm Pois} \rangle \equiv \sum_{i=0}^\infty 
\rme^{-\bar{n}_0} \frac{\bar{n}_0^i}{i!} | i \rangle
= \rme^{\bar{n}_0 ( \hat{a}^\dagger - 1)} | 0 \rangle\, .
\label{init}
\end{eqnarray}
For the moment, we focus on the case 
\begin{equation}
\alpha \bar n_0 -\frac{\beta}{\Omega}\bar n_0^2 >0\, ,
\label{n0con}
\end{equation}
and the other case will be  discussed at the end of this section.
By inserting the identity \eref{identity_eq}
into (\ref{eq_observables}) with (\ref{psit}) and (\ref{init}),
we express $\bra A(n(\tau)) \ket$ as 
\begin{eqnarray}
\fl \qquad
\bra A(n(\tau))\ket =& \lim_{\Delta t \to 0}
\left( \prod_{t=0}^\tau 
\int_{-\infty}^\infty \rmd \phi_{t} 
\int_{-\infty}^\infty \frac{\rmd \varphi_{t}}{2\pi}
\right)
\langle \mathcal{P} | {\cal A}(\hat{a}) | \phi_{t} \rangle 
\nonumber \\
&\times \left[ 
\prod_{t = \Delta t}^{\tau} 
\langle -\rmi \varphi_t | \rme^{-\hat{H} \Delta t} | \phi_{t-\Delta t} 
\rangle
\rme^{-\rmi \phi_t \varphi_t}
\right] 
\rme^{-\rmi \varphi_0 \phi_0}
\langle -\rmi \varphi_0 | \bar{n}_0 ; {\rm Pois}\rangle\, ,
\label{Ant} 
\end{eqnarray}
where the time index $t$ runs from zero to $\tau$ in steps of $\Delta t$.
(We assume $\tau=N \Delta t$ with integer $N$.) 
Since each term of $\hat{H}$ in 
\eref{time_evolution_operator} is arranged in normal order,
we derive 
\begin{eqnarray}
\langle -\rmi \varphi_t |  \rme^{-\hat{H} \Delta t} | 
\phi_{t-\Delta t} \rangle
= \langle -\rmi \varphi_t | \phi_{t - \Delta t} \rangle
\exp(- H(\rmi \varphi_t, \phi_{t - \Delta t}) \Delta t)\, ,
\label{eq_replacement_with_coherent_state_variables}
\end{eqnarray}
where $H(\rmi \varphi_t, \phi_{t - \Delta t})$
is defined by simply replacing $\hat{a}^\dagger$ and 
$\hat{a}$ in $\hat H$ with $\rmi \varphi_t$ and 
$\phi_{t - \Delta t}$, respectively. 
We also have
$
\langle -\rmi \varphi_t | \phi_{t - \Delta t} \rangle 
= \exp(\rmi \varphi_t \phi_{t - \Delta t})\, 
$,
$
\langle \mathcal{P} | \mathcal{A}(\hat{a})| 
\phi_\tau \rangle 
= \langle 1 | \phi_\tau \rangle \mathcal{A}(\phi_\tau)
= \mathcal{A}(\phi_\tau) \rme^{\phi_\tau}\, 
$,
and 
$
\langle -\rmi \varphi_0 | \bar{n}_0 ; {\rm Pois} \rangle
= \exp\left( \rmi \varphi_0 \bar{n}_0 - \bar{n}_0 \right)\, .
$
By substituting these results into (\ref{Ant}),
we obtain a path-integral expression  
\begin{eqnarray}
\fl \qquad
\bra A(n(\tau)) \ket
= \lim_{\Delta t \to 0} 
\left( \prod_{t=0}^\tau \int_{-\infty}^\infty \rmd \phi_{t} 
\int_{-\infty}^\infty \frac{\rmd \varphi_{t}}{2\pi}\right)
\mathcal{A}(\phi_\tau) \exp\left[ 
-S( \{ \rmi \varphi \}, \{\phi\})
\right]\, ,
\label{path_integral_eq}
\end{eqnarray}
where the ``action'' $S$ is calculated as
\begin{eqnarray}
\fl \quad
&&S(\{ \rmi \varphi\}, \{\phi\})
= - \phi_\tau + \rmi \varphi_0 (\phi_0-\bar{n}_0)+ \bar{n}_0 
\nonumber \\
\fl
& &\quad + \sum_{t = \Delta t}^\tau \Delta t \left[\rmi \varphi_t 
\frac{\phi_t - \phi_{t-\Delta t}}{\Delta t}
- \alpha \left( (\rmi \varphi_t)^2 - \rmi \varphi_t \right) 
\phi_{t - \Delta t} 
- \frac{\beta}{\Omega} 
\left(\rmi \varphi_t - (\rmi \varphi_t)^2\right) \phi_{t - \Delta t}^2
\right]\, . 
\label{action}
\end{eqnarray}
% Langevin description 
Although $S$ is a complex valued functional as it is, 
the integration 
with respect to $\varphi_t$ in (\ref{path_integral_eq})
yields a  real valued functional of $\phi$. 
Here, the integration of $\varphi_t$ in 
(\ref{path_integral_eq}) is carried out 
from $-\infty$ to $+\infty$ along  the real axis. 
We assume that 
the integration path can be shifted to the 
straight line from $-\infty-\rmi$ to $+\infty-\rmi$ with 
keeping the integration value unchanged. This shift
is equivalent to the replacement of 
$\varphi_t$ with $\bar \varphi_t-\rmi$, where $\bar \varphi_t$ 
is a real variable. Then, the action (\ref{action})
becomes  
\begin{eqnarray}
\fl \quad
&&\tilde S(\{\rmi \bar{\varphi}\}, \{\phi\})
= \rmi \bar{\varphi}_0 (\phi_0-\bar{n}_0)  \nonumber \\
\fl
&&\quad + \sum_{t = \Delta t}^\tau \Delta t \left[\rmi \bar{\varphi}_t 
\frac{\phi_t - \phi_{t-\Delta t}}{\Delta t}
- \alpha \left( (\rmi \bar{\varphi}_t)^2 + \rmi \bar{\varphi}_t\right) 
\phi_{t - \Delta t} 
+ \frac{\beta}{\Omega} 
\left(\rmi \bar{\varphi}_t + (\rmi \bar{\varphi}_t)^2\right) 
\phi_{t - \Delta t}^2
\right]\, . 
\label{tildeaction}
\end{eqnarray}
Next, we linearize the terms quadratic in $\bar \varphi_t$
by introducing an auxiliary real variable $y_t$:
\begin{eqnarray}
\fl \quad
\exp\left[ 
 (\rmi \bar{\varphi_t})^2 
\left( \alpha \phi_{t-\Delta t} 
- (\beta/\Omega) \phi_{t-\Delta t}^2 \right)(\Delta t) \right]&&
\nonumber \\
&&\hspace{-65mm}
= \int_{-\infty}^\infty \frac{\rmd y_t}{\sqrt{2\pi}}\, 
\exp\left[
- \frac{1}{2} y_t^2 + \rmi \bar{\varphi_t} 
\sqrt{2\left(\alpha \phi_{t-\Delta t} 
- (\beta/\Omega)\phi_{t-\Delta t}^2\right)}\, (\Delta t)^{1/2}\, y_t
\right]\, .
\label{HS}
\end{eqnarray}
Note that this identity is valid only when
\begin{equation}
\alpha \phi_{t-\Delta t} - (\beta/\Omega)\phi_{t-\Delta t}^2
\ge 0\, . 
\label{pcon}
\end{equation}
We conjecture that trajectories that do not satisfy this condition
do not contribute the path-integration in 
(\ref{path_integral_eq}) under the condition
(\ref{n0con}). 
See a discussion  below (\ref{DP:final}). 
Finally, by employing the Fourier transformation formula of 
Dirac's delta function, we can rewrite (\ref{path_integral_eq}) as 
\begin{eqnarray}
\fl
\bra A(n(\tau)) \ket\! =\! \lim_{\Delta t \to 0} \!
\left( \prod_{t=\Delta t}^\tau  
\int_{-\infty}^\infty \! \rmd \phi_{t} \!
\int_{-\infty}^\infty \! \frac{\rmd y_{t}}{\sqrt{2\pi}}
{\rm e}^{- \frac{1}{2} y_t^2}\! \right )\!
\mathcal{A}(\phi_\tau) \!
\left[\prod_{t=\Delta t}^\tau \! \delta\left( 
\Phi(\phi_t,\phi_{t-\Delta t},y_t)  \right) 
\right]_{\phi_0=\bar n_0}\! ,
\label{path_integral_final}
\end{eqnarray}
where
\begin{eqnarray}
\Phi(\phi_t,\phi_{t-\Delta t},y_t)
&=& 
\phi_t - \phi_{t-\Delta t}
-\left( \alpha \phi_{t - \Delta t} 
- \frac{\beta}{\Omega} \phi_{t - \Delta t}^2 
\right) (\Delta t)
\nonumber \\
& & -\sqrt{2 \left( \alpha \phi_{t - \Delta t} 
- \frac{\beta}{\Omega} \phi_{t - \Delta t}^2 \right)}
(\Delta t)^{1/2}y_t\, .
\label{Phidef}
\end{eqnarray}
The equation $\Phi=0$ in the limit $\Delta t \to 0$ 
is expressed as 
\begin{eqnarray}
\frac{\rmd}{\rmd t} \phi = \alpha \phi - \frac{\beta}{\Omega} \phi^2
+  \sqrt{2 \left(\alpha \phi - \frac{\beta}{\Omega}\phi^2 \right)} \,
\cdot\eta \, ,
\label{eq_sFKPP_Doi_Peliti}
\end{eqnarray}
where the quantity $\eta(t)$ defined as the continuum limit
of $y_t/\sqrt{\Delta t}$ is a white Gaussian noise that satisfies
$\bra \eta(t) \eta(t') \ket=\delta(t-t')$, and  
the symbol $\cdot$ in front of $\eta$ represents the 
product in the sense of It\^o.  Therefore, 
the formula (\ref{path_integral_final}) with (\ref{Phidef})
is written as 
\begin{equation}
\bra A(n(\tau)) \ket = \bra \mathcal{A} (\phi(\tau)) \ket_\phi\, ,
\label{DP:final}
\end{equation}
where $\bra \ \ket_\phi$ represents the expectation
value over  noise sequences of the Langevin equation
(\ref{eq_sFKPP_Doi_Peliti}). It should be noted that 
trajectories  $\phi(t)$ generated by (\ref{eq_sFKPP_Doi_Peliti})
with the initial condition $\phi(0)=\bar n_0$ satisfy 
%the condition 
$\alpha \phi-(\beta/\Omega) \phi^2 \ge 0$ for any time $t$
as discussed in Ref.~\cite{Pechenik1999}.
Then, one may obtain 
(\ref{tildeaction}) starting from (\ref{eq_sFKPP_Doi_Peliti}). 
This implies that the conjecture mentioned  below (\ref{pcon}) 
is valid.

% s-FKPP

We choose $A(n)=n/\Omega$ in (\ref{DP:final}) as the simplest
example. By setting $u(t)=\phi(t)/\Omega$, we have
\begin{equation}
\bra \frac{n(\tau)}{\Omega} \ket =\bra u(\tau) \ket_u\, ,
\label{simple_case}
\end{equation}
where $\bra \ \ket_u$ represents the expectation
value over  noise sequences of the Langevin equation
\begin{eqnarray}
\frac{\rmd}{\rmd t} u = \alpha u - \beta u^2
+  \sqrt{\frac{2 (\alpha u - \beta u^2) }{\Omega}} \,
\cdot\eta \, .
\label{eq_sFKPP_Doi_Peliti-rho}
\end{eqnarray}
Surprisingly enough, the average behavior of the density 
$n/\Omega$ is exactly described by the Langevin equation 
(\ref{eq_sFKPP_Doi_Peliti-rho}) for any $\Omega$.
The Langevin equation (\ref{eq_sFKPP_Doi_Peliti-rho}) 
supplemented with a diffusion term is called the stochastic Fisher 
Kolmogorov-Petrovsky-Piscounov (s-FKPP) equation,
which  appears in several research fields \cite{Panja2004}
(for example, see Ref.~\cite{Munier2009} for applications 
to the problems of high-energy hadron scattering). 
Historically, the s-FKPP equation for the chemical reaction 
$X \rightleftarrows X+X$ was derived on the basis of the 
coherent-state path-integral expression with the complex 
parameterization $(z,z^*)$ \cite{Pechenik1999} in
studying  non-trivial noise effects for front-propagation 
\cite{Brunet1997,Brunet2006}. Here, it is worthwhile
noting that $\phi$ does {\it not} correspond to the number of particles, 
as we can see the difference between the two polynomials 
$A$ and $\mathcal{A}$ that appear in (\ref{DP:final}).
It is only when we take the simplest case $A(x)=x$ 
discussed in (\ref{simple_case}) that ${\mathcal A}(x)$ 
becomes equivalent to $A(x)$.
Therefore, one must {\it not} interpret that 
(\ref{eq_sFKPP_Doi_Peliti-rho}) is the equation for a 
fluctuating density. Indeed, (\ref{eq_sFKPP_Doi_Peliti-rho}) is different
from that obtained by system-size expansion methods for the
master equation (\ref{master_eq}). (Compare (\ref{eq_sFKPP_Doi_Peliti-rho}) 
with (\ref{Lan}) derived in the next section.) 

% technical remarks 

At the end of this section,  we address three technical remarks.
First, we discuss the validity of the assumption mentioned 
above (\ref{tildeaction}). We do not have a  mathematical
proof for the claim that the integration path can be shifted
as such described there. However, this is plausible because 
we can prove the result  (\ref{DP:final}) in the following manner:
It is sufficient to consider the case where $A(x)$ is 
given by $A_k(x)=x(x-1)\cdots (x-k+1)$ with arbitrary 
positive integer $k$. The corresponding polynomial $\mathcal{A}(x)$
becomes $\mathcal{A}_k(x)=x^k$. By taking the $k$-th derivative of 
the identity 
\begin{equation}
\sum_{n=0}^\infty \frac{s^n \bar n_0^n}{n!}={\rm e}^{\bar n_0 s}
\end{equation}
with respect to $s$, and setting $s=1$, we obtain
\begin{equation}
\sum_{n=0}^\infty \frac{\bar n_0^n \rme^{-\bar n_0}}{n!}
n(n-1)\cdots (n-k+1)=\bar n_0^k \, ,
\end{equation}
which means 
\begin{equation}
\bra A_k(n(0)) \ket=\mathcal{A}_k(\bar n_0)\, .
\label{init-ident}
\end{equation}
Next, following the It\^o formula, we find from the 
Langevin equation (\ref{eq_sFKPP_Doi_Peliti}) for $\phi$
\begin{eqnarray}
\frac{\rmd \bra \mathcal{A}_k(\phi) \ket_\phi}{\rmd t}
&= & 
\bra \left(\alpha \phi-\frac{\beta}{\Omega}\phi^2\right) 
\der{\mathcal{A}_k(\phi)}{\phi} \ket_{\phi}
+  
\bra \left(\alpha \phi-\frac{\beta}{\Omega}\phi^2\right)
\tder{\mathcal{A}_k(\phi)}{\phi} \ket_{\phi}
\nonumber \\
&=&
\alpha 
\bra k \mathcal{A}_k +k(k-1) \mathcal{A}_{k-1} \ket_{\phi}
-\frac{\beta}{\Omega}
\bra k \mathcal{A}_{k+1} +k(k-1) \mathcal{A}_{k} \ket_{\phi}.
\label{dAdt}
\end{eqnarray}
On the other hand, by using the master equation (\ref{master_eq}), 
we can confirm that the differential equation for 
$\rmd \bra A_k (n(t)) \ket/\rmd t$ has the structure 
identical with that of the right-hand side of (\ref{dAdt}). 
This observation 
combined with (\ref{init-ident}) leads to $\bra A_k(n(t))\ket
=\bra \mathcal{A}_k(\phi(t)) \ket_\phi$ for any $k$ at 
any $t$. 
Since arbitrary polynomials can be expressed as a linear
combination of  $\{A_k \}$, (\ref{DP:final}) has been proved.

% second and third remark

Second, when $\bar n_0$ does not satisfy the condition (\ref{n0con}), 
the integration formula (\ref{HS}) is not available. 
Therefore, the result (\ref{DP:final})  with the Langevin equation 
(\ref{eq_sFKPP_Doi_Peliti}) is valid only for the case (\ref{n0con}). 

Finally, let us consider two-time quantities such as 
$\bra A(n(t_1)) A(n(t_2)) \ket$. In order to calculate such 
a quantity, we need a path-integral expression of the conditional 
probability $P(n,\tau|n_0,0)$. If we replace the Poisson 
initial condition with $\delta(n-n_0)$, the path-integral
expression in this section becomes complicated. 

%%%%%%%%%%%%%%%%%%%%%%%%%%%%%%%%%%%%%%%%%%%%%%%%%%%%%%%
\section{Density fluctuations} \label{sec:new}         %
%%%%%%%%%%%%%%%%%%%%%%%%%%%%%%%%%%%%%%%%%%%%%%%%%%%%%%%

% so-called Cole-Hopf transformation 

In order to have a direct correspondence with density
fluctuations in the coherent-state path-integral expression, 
a non-linear transformation in the action 
(\ref{action}) 
\begin{eqnarray}
\phi = 
\nu \, \rme^{- \chi}\, , 
\quad 
\rmi \varphi = \rme^{\chi}
\label{CH}
\end{eqnarray}
has been employed \cite{Tauber2005,Janssen2001,Frederic2001,Andreanov2006}, 
which is called the Cole-Hopf transformation. 
(In the standard formulation with  $(z,z^*)$,  $(\phi, \rmi \varphi )$ in 
(\ref{CH}) is simply replaced with $(z,z^*)$.) Quite formally, 
substituting (\ref{CH}) into (\ref{action}), neglecting terms 
with third and higher powers of $\chi$, expecting cancellation 
of several terms appearing in contributions at $t=\tau$, 
and avoiding  considerations on the integration path, one can 
obtain the dynamical action of a  Langevin equation for the 
variable $\nu/\Omega$. This Langevin equation coincides with 
that  obtained by system-size expansion methods.

% our motivation 

Although the final result is plausible, it seems difficult to judge 
the validity of the procedures. In particular, in the standard
formulation with $(z,z^*)$, there is no complex number $\chi$
that would satisfy  $z = \nu \, \rme^{- \chi}$ and $z^* = \rme^{\chi}$,
as easily checked for an example $z=1+\rm i$. In the formulation 
with $(\phi,\rmi \varphi)$, the integration path of $\chi$ is 
described by $\chi=\log |\varphi|+\rmi \pi {\rm sgn}(\varphi) /2$ with
$-\infty < \varphi <0$ and $0< \varphi < \infty$. The calculation
after that seems complicated. Toward a justification of (\ref{CH}), 
recently, an operator version of the Cole-Hopf transformation has been 
presented \cite{Lefevre2007}; 
but its mathematical foundation is  not obvious.

% Our idea

Based on these understandings, we propose a framework in which the 
calculation procedures mentioned above are formulated without 
any mathematical difficulties.  
Our basic idea is to introduce a  new decomposition of unity 
associated with the Cole-Hopf transformation. As a preparation,
motivated by (\ref{CH}), we define
\begin{eqnarray}
\tilde{z } \equiv z/|z|
\end{eqnarray}
for any non-zero complex number $z$. In the argument below, 
the complex variable $z$ is always connected to  two real 
variables $\mu$ and $\theta$ as  
\begin{equation}
z = \mu \, \rme^{-\rmi \theta}\, ,
\end{equation}
where $\mu \ge 0$ and $ -\pi \le \theta \le \pi$. 
Then, $\tilde{z} =\rme^{-\rmi \theta}$. By using coherent states
parametrized by $z$ and $\tilde z$, we find a new 
decomposition of unity as follows:
\begin{eqnarray}
\bm{1}
&=& \sum_\ell \frac{1}{\ell!} | \ell \rangle \langle \ell |  \nonumber \\
&=&
\sum_{\ell} \frac{1}{\ell! \ell!}
\int_{0}^\infty \rmd \mu \,
\rme^{-\mu} \mu^\ell | \ell \rangle \langle \ell | \nonumber \\
&=&
\sum_{\ell,m} \frac{1}{\ell! m!}
\int_{0}^\infty \rmd \mu \int_{-\pi}^\pi \frac{\rmd \theta}{2\pi}
\rme^{-\mu} \mu^\ell \rme^{- \rmi (\ell-m) \theta} 
| \ell \rangle \langle m | \nonumber \\
&=&
\int_{0}^\infty \rmd \mu \int_{-\pi}^\pi \frac{\rmd \theta}{2\pi}
| z \rangle \langle \tilde{z} | \rme^{- \mu}\, .
\label{new-ident}
\end{eqnarray}
We shall employ this decomposition of unity 
for constructing a path-integral expression.

% exact expression 

We consider the transition probability $P(\nf,\tau| \nint,0)$, which 
is the probability of finding $\nf$ particles at time $t=\tau$ 
provided that there are $\nint$ particles at $t=0$. This is 
expressed as
\begin{equation}
P(\nf,\tau|\nint,0)=\frac{1}{\nf !} 
\bra \nf \left\vert \rme^{-\hat H \tau} \right\vert \nint \ket\, .
\label{transition_prob}
\end{equation}
Its path-integral expression on the basis of (\ref{new-ident}) is 
written as 
\begin{eqnarray}
P(\nf,\tau|\nint,0)
&=& \lim_{\Delta t \to 0}
\left(\prod_{t=0}^\tau   \int_0^\infty \rmd \mu_{t}
\int_{-\pi}^{\pi} \frac{ \rmd \theta_{t}}{2\pi}\right)
\frac{1}{\nf !} \bra \nf |z_\tau \ket 
\nonumber \\
&&\times
\left[ 
\prod_{t = \Delta t}^{\tau} 
\langle \tilde{z}_t | 
\rme^{-\hat{H} \Delta t} | z_{t-\Delta t} \rangle
\rme^{- \mu_t}
\right] 
\rme^{- \mu_0}
\langle \tilde{z}_0 | \nint \rangle \, . 
\label{eq_path_integral_for_new_coherent_state}
\end{eqnarray}
(See \ref{path-particle} for the corresponding
path-integral expression in terms of particle numbers 
instead of coherent states.)
As did in the previous section,  we have
$
\langle \tilde{z}_t |  \rme^{-\hat{H} \Delta t} | 
z_{t-\Delta t} \rangle
= \langle \tilde {z}_t | z_{t - \Delta t} \rangle
\exp(- H(\tilde{z}_t^*, z_{t - \Delta t}) \Delta t),
$
where $H$ is expressed as 
\begin{eqnarray}
H(\tilde{z}_t^*, z_{t - \Delta t})
=
&- \alpha \left(
\left(\rme^{\rmi \theta_t} \right)^2 - \rme^{\rmi \theta_t}
\right)
\mu_{t-\Delta t}\, \rme^{-\rmi \theta_{t-\Delta t}} \nonumber \\
&- \frac{\beta}{\Omega}
\left(
\rme^{\rmi \theta_t} - \left(\rme^{\rmi \theta_t} \right)^2 
\right)
\left(\mu_{t-\Delta t} \right)^2 \rme^{-2 \rmi \theta_{t-\Delta t}}\, .
\end{eqnarray}
We then notice 
\begin{eqnarray}
%\fl
\prod_{t = \Delta t}^{\tau} 
\langle \tilde{z}_t | z_{t - \Delta t} \rangle
\rme^{-\mu_t}
&=& 
\prod_{t = \Delta t}^{\tau} 
\exp\left( -\mu_t+\mu_{t-\Delta t} \, 
\rme^{\rmi (\theta_t - \theta_{t - \Delta t})}
\right) 
\nonumber \\
\fl
&=& 
\prod_{t = \Delta t}^{\tau} 
\rme^{ -\mu_t+\mu_{t-\Delta t} 
+\rmi \mu_{t-\Delta t} 
(\theta_t - \theta_{t - \Delta t})+O(\Delta t^2)}
\nonumber \\
\fl
&=&
\rme^{-\mu_\tau+\mu_0+\rmi \theta_\tau\mu_\tau-\rmi\theta_0 \mu_0}
\prod_{t = \Delta t}^{\tau} 
\rme^{ -\rmi \theta_{t} 
({\mu}_t - {\mu}_{t - \Delta t})+O(\Delta t^2)}\, ,
\label{eq_chi_braket}
\end{eqnarray}
where we have used the estimation
$\theta_t-\theta_{t-\Delta} \simeq O(\Delta t)$.
Furthermore, we have 
$
\bra \nf | z_\tau \ket
= 
z_\tau^{\nf}
=
\mu_\tau^{\nf}\rme^{-\rmi \nf \theta_\tau}
$
and 
$
\langle \tilde{z}_0 | \nint \rangle
= \rme^{\rmi \theta_0 \nint }.
$
Substitution of these results into 
\eref{eq_path_integral_for_new_coherent_state} yields
\begin{eqnarray}
%\fl 
P(\nf,\tau|\nint,0)
= 
& &
\lim_{\Delta t \to 0}
\left(\prod_{t=0}^\tau   \int_0^\infty \rmd \mu_{t}
\int_{-\pi}^{\pi} \frac{ \rmd \theta_{t}}{2\pi}\right)
\rme^{-\rmi \theta_0(\mu_0-\nint)+\rmi 
\theta_\tau(\mu_\tau-\nf)} \nonumber \\ 
& & 
\times \frac{1}{\nf !} \mu_\tau^{\nf} \rme^{-\mu_\tau} 
\exp[-\mathcal{S}(\{\rmi \theta_t\}, \{ \mu_t \} )]\, ,
\label{path:mu}
\end{eqnarray}
with 
\begin{eqnarray}
\mathcal{S}(\{\rmi \theta_t\}, \{ \mu_t \} )
=\sum_{t=\Delta t}^\tau[
H(\tilde z_t^*, z_{t-\Delta t} )\Delta t
+\rmi \theta_t(\mu_t-\mu_{t-\Delta t})]\, .
\label{action:mu}
\end{eqnarray}
Here, we perform the integrations with respect to
$\theta_0$ and $\theta_\tau$. Noting that 
the limit $\Delta t \to 0 $ is taken in the final 
expression, we obtain
\begin{eqnarray}
\fl \quad
P(\nf,\tau|\nint,0)
= \lim_{\Delta t \to 0}
\left(\prod_{t=\Delta t}^{\tau-\Delta t} 
\int_0^\infty \rmd \mu_{t}
\int_{-\pi}^{\pi} \frac{ \rmd \theta_{t}}{2\pi}\right)
\left.
\frac{1}{\nf !} \nf^{\nf}\rme^{-\nf}
\rme^{-\mathcal{S}(\{\rmi \theta_t\}, \{ \mu_t \} )}
\right\vert_{\mu_0=\nint, \mu_\tau=\nf} .
\label{path:mu-final}
\end{eqnarray}
This expression is exact and if one simply replaces 
${\rm i}\theta_t$ and $\mu_t$ with $\chi_t$ and $\nu_t$ 
respectively,
the resulting action $\mathcal{S}(\{ \chi_t \}, \{ \nu_t \} )$ 
is equal to the one obtained by a formal procedure with the 
Cole-Hopf transformation (\ref{CH}).  Therefore, we
claim that our argument provides a mathematical
foundation for the Cole-Hopf transformation.

% address the question

Now, recalling  that $\Omega $ is a dimensionless 
volume of the system, we focus on  the regime 
$\nf \gg 1$ and $\nint \gg 1$ under the assumption
$\Omega \gg 1$. 
More explicitly, by setting  $\rhoi=\nint/\Omega $ 
and $\rhof=\nf/\Omega $, we assume a large deviation 
property
\begin{equation}
P(\rhof \Omega,\tau| \rhoi \Omega ,0)
\simeq \rme^{-\Omega \mathcal{F}(\rhof,\tau|\rhoi,0)}\, ,
\label{LD}
\end{equation}
where $\rhoi$ and $\rhof $ are the particle densities 
at initial and final times and are finite in general.
Note that the relation $A(\Omega) \simeq B(\Omega)$ in 
this paper means $(\log A(\Omega)-\log B(\Omega))/\Omega \to 0$
in the limit $\Omega \to \infty$. The quantity 
$\mathcal{F}(\rhof,\tau|\rhoi,0)$, which is 
called a large deviation function, characterizes 
fluctuation properties of a density in a macroscopic 
system. The problem we consider here is to derive 
a simpler stochastic system that reproduces 
$\mathcal{F}(\rhof,\tau|\rhoi,0)$ defined in (\ref{LD}).

% calculation of the large deviation

Since $\mu_0$ and $\mu_\tau$ are fixed to $\rhoi \Omega$ 
and $\rhof \Omega$ respectively, the dominant contribution 
of the $\mu_t$ integration in (\ref{path:mu-final}) 
comes from a region $\mu_t \sim O(\Omega)$. On the other
hand, since there is the term 
$\exp (\rmi \theta_t (\mu_t-\mu_{t-\Delta t}))$ 
in the integrand, the dominant contribution of the $\theta_t$ 
integration comes from a region $\theta_t \sim O(1/\Omega)$. 
From these observations, we make the transformation of variables 
as $\rho_t = \mu_t/\Omega$ and $\pi_t = \Omega \theta_t$ 
in the path-integral expression (\ref{path:mu-final}) so that
the integration over a region where 
$\rho_t \simeq O(1)$ and $\pi_t \simeq O(1)$
provides the dominant contribution. We also note the relation
$\nf! \simeq \nf^{\nf}\rme^{-\nf}$ valid for $\nf \gg 1$.
We then have 
\begin{eqnarray}
\fl \quad
\rme^{-\Omega \mathcal{F}(\rhof,\tau|\rhoi,0)}
\simeq  \lim_{\Delta t \to 0}
\left(\prod_{t=\Delta t}^{\tau-\Delta t} 
\int_0^\infty \rmd \rho_{t}
\int_{-\infty}^{\infty} 
\frac{ \rmd \pi_{t}}{2\pi}\right)
\left.
\rme^{- \bar \mathcal{S}
(\{\rmi \pi_t \}, \{  \rho_t \} )
}\right\vert_{\rho_0=\rhoi, \rho_\tau=\rhof}\, ,
\label{path:rho}
\end{eqnarray}
where 
$\bar \mathcal{S}(\{\rmi \pi_t\}, \{ \rho_t \} )$
is expanded in terms of $1/\Omega$ and $\Delta t$
as 
\begin{eqnarray}
\fl \quad
\bar \mathcal{S}(\{\rmi \pi_t\}, \{ \rho_t \} )
= (\Delta t) \sum_{t=\Delta t}^\tau
\left[ 
\rmi \pi_t \Psi_1(\rho_t,\rho_{t-\Delta t})
+\frac{1}{2\Omega} (\pi_t)^2 \Psi_2(\rho_{t-\Delta t})
+R(\rmi \pi_t, \rho_{t-\Delta t}) \right]
\label{Sbar}
\end{eqnarray}
with
\begin{eqnarray}
\Psi_1(\rho_t,\rho_{t-\Delta t})
&=& \frac{\rho_t-\rho_{t-\Delta t}}{\Delta t}
-\left(\alpha \rho_{t-\Delta t}-\beta \rho_{t-\Delta t}^2\right)\, ,
\label{psi1}
\\
\Psi_2(\rho_{t-\Delta t})
&=& \alpha \rho_{t-\Delta t}+\beta \rho_{t-\Delta t}^2\, ,
\label{psi2}
\\
R(\rmi \pi_t, \rho_{t-\Delta t})
&=& O\left(\frac{1}{\Omega^2}\right)+O(\Delta t)\, .
\end{eqnarray}
The $\pi_t $ integration in (\ref{path:rho}) with 
$R(\rmi \pi_t, \rho_{t-\Delta t})$ in $\bar \mathcal{S}$ 
being ignored yields the formula
\begin{equation}
\mathcal{F}(\rhof,\tau|\rhoi,0)
= 
\lim_{\Delta t \to 0}
\min_{\rhoi\to \rhof}
(\Delta t) \sum_{t=\Delta t}^\tau
\frac{[\Psi_1(\rho_t,\rho_{t-\Delta t})]^2}{2\Psi_2(\rho_{t-\Delta t})}\, ,
\label{LD:formula}
\end{equation}
where $\min_{\rhoi \to \rhof}$ represents the minimization with
respect to $\{ \rho_t \}$ with the boundary conditions 
$\rho_0=\rhoi$ and $\rho_\tau=\rhof$ fixed.
If one includes the term $R(\rmi \pi_t, \rho_{t-\Delta t})$ 
in $\bar \mathcal{S}$ and treats it as a perturbation,
one can show that it does not contribute to the large 
deviation function $\mathcal{F}$. Therefore, as far as 
we are concerned with the large deviation property, we 
may neglect $R(\rmi \pi_t, \rho_{t-\Delta t})$ in $\bar \mathcal{S}$.

% result 

Then, by applying the same procedures as (\ref{HS}) and 
(\ref{path_integral_final}) in the previous section
to the expression (\ref{path:rho}) and (\ref{Sbar}), 
we arrive at the final result 
\begin{eqnarray}
P(\rhof \Omega , \tau |\rhoi \Omega,0) 
\simeq  \bra \delta(\rho(\tau)-\rhof) \ket_{\rho}\, ,
\label{result-LD}
\end{eqnarray} 
where $\bra \ \ket_\rho$ represents the average 
over trajectories given by the Langevin equation
\begin{eqnarray}
\frac{\rmd}{\rmd t} \rho = \alpha \rho - \beta \rho^2
+  \sqrt{ \frac{\alpha \rho + \beta \rho^2}{\Omega} } \, \cdot \eta
\label{Lan}
\end{eqnarray}
under the initial condition $\rho(0)=\rhoi$. 
This result coincides with the result obtained by any system-size 
expansion methods for the master equation. 
Since (\ref{result-LD}) holds for any 
$\tau$, the variable $\rho$ defined by $\mu/\Omega$ corresponds to 
the density in this description.

%%%%%%%%%%%%%%%%%%%%%%%%%%%%%%%%%%%%%%%%%%%%%%%%
\section{Effects of diffusion} \label{sec:diff} %
%%%%%%%%%%%%%%%%%%%%%%%%%%%%%%%%%%%%%%%%%%%%%%%%

% motivation

In the main part of this paper, we have focused 
on the case where diffusion processes 
are sufficiently fast. This formulation can be 
extended to cases with spatially heterogeneous 
fluctuations. The simplest example in such systems 
is given by a one-dimensional lattice model, 
which we shall explain. 

% model

Let $\Lambda =\{ i| i=0, 1, \cdots, L\}$ be a one-dimensional lattice. 
The reaction  $X \rightleftarrows X+X$ occurs on each site in the 
lattice, and a particle moves to a nearest neighbor site at rate $d$. 
The state of the system is specified by a set of particle 
numbers on each site, $\bv{n}=(n_1,\cdots,n_L)$. 
Boundary conditions are assumed appropriately,
depending on the situation we consider. We then write the
master equation for $P(\bv{n},t)$ and transform it to the equation
for a single vector. The time-evolution operator $\hat H$
in the equation for this vector is given by 
\begin{eqnarray}
\fl \quad
\hat{H}= 
d \sum_{\langle ij \rangle \in {\mathcal{B}}} 
(\hat{a}_i^\dagger - \hat{a}_j^\dagger) (\hat{a}_i - \hat{a}_j)
+ \sum_{i \in \Lambda} \left[
- \alpha  (\hat{a}_i^\dagger - 1) \hat{a}_i^\dagger \hat{a}_i 
- \beta(1 - \hat{a}_i^\dagger) \hat{a}_i^\dagger \hat{a}_i^2
\right]\, ,
\label{time_evolution_operator_diffusion}
\end{eqnarray}
where $\mathcal{B}$ represents a set of all the nearest 
neighbor pairs,
and $\hat a_i^\dagger $ and $\hat a_i$ are introduced on each site
in a  manner similar to $\hat a^\dagger $ and $\hat a$ in section 
\ref{sec:model}. 

% address  a  question /general

Now, as a straightforward extension of the formulation
discussed in section 
\ref{sec:new}, we define a real variable $\mu_i(t)$ which can be 
identified with the particle number on a site $i$ at time $t$. 
Our
central question for this model is to derive an {\it effective} 
stochastic system which is defined as the simplest description 
reproducing large-distance and long-time behavior.
More explicitly, 
the effective stochastic system might be defined by a Langevin
equation which provides a large deviation functional for 
the transition probability of a density field. Since the large deviation
functional is related to an effective action in the field-theoretical 
language \cite{Eyink1996,Hochberg1999}, the question is equivalent
to an identification of the fixed point in renormalization group 
flow. However, as far as we know, a concrete calculation based 
on such a formulation has not been reported yet.

% conventional description

Putting aside developing such a theoretical framework, we here 
present a conventional derivation by considering a special situation.
We first assume that one-site corresponds to a coarse-grained cell 
in which diffusion processes are sufficiently fast so that chemical 
components can be thought to be uniform in the cell. 
Let $\Delta x$ be a dimensionless physical size of the cell 
by setting a reaction length as unity (it thus corresponds to 
$\Omega$ in the previous sections).  As discussed in the first 
paragraph of  section \ref{sec:model}, we then replace $\beta$
in (\ref{time_evolution_operator_diffusion}) with 
$\tilde \beta/\Delta x$ under the assumption $\Delta x \gg 1 $.
We also assume that there exists a density field $\rho(x,t)$,
$0 \le x \le L\Delta x$, that satisfies 
$\rho(i \Delta x,t)=\mu_i(t)/(\Delta x)$ and  
$\sup_{x} |\partial_x^k \rho(x,t)| (\Delta x)^k = O(\epsilon^k)$
with a small positive constant $\epsilon$. 
The latter condition means that the density 
field is slowly varying in space. Since the longest wavelength 
of fluctuations is $O(L\Delta x)$, $\epsilon$ corresponds to $1/L$.
We thus consider the case $ L \gg 1$. By repeating the argument 
in section \ref{sec:new} with the replacement of $\Omega$ with 
$\Delta x$,  
one may derive
\begin{eqnarray}
P(\bv{n}^{\rm fin}, \tau |\bv{n}^{\rm init},0) 
\simeq  \bra \prod_{i=1}^L
\delta
\left(\rho(i \Delta x,\tau)- \frac{{n}^{\rm fin}_i}{\Delta x}
\right) 
\ket_{\rho}\, ,
\label{result-LD-space}
\end{eqnarray} 
where $\simeq $ represents the equality valid asymptotically
in the limit $\Delta x \gg 1$, and $\bra \ \ket_\rho$ 
represents the average over trajectories of a space-discretized form 
of the spatially extended Langevin equation 
\begin{eqnarray}
\pder{\rho}{t} = \alpha \rho - \tilde \beta \rho^2+\tilde d  \partial_x^2\rho
+\sqrt{\alpha \rho + \tilde \beta \rho^2} \cdot \xi_1
+\partial_x \left(\sqrt{2 \tilde d \rho} \cdot \xi_2 \right)\, ,
\label{Lan-space}
\end{eqnarray}
under the initial condition $\rho(i\Delta x,0)={n}^{\rm init}_i/\Delta x$.
Here, we defined $\tilde d \equiv d (\Delta x)^2$ and 
$\xi_1$ and $\xi_2$ as noises satisfying
\begin{equation}
\bra \xi_i(x,t) \xi_j(x',t')\ket=\delta_{i,j}\delta(x-x')\delta(t-t')\, .
\label{noise-sp}
\end{equation}
We note that the discretized form of (\ref{noise-sp}) is 
\begin{equation}
\bra \xi_i(\ell \Delta x,t) \xi_j(\ell'\Delta x,t')
\ket=\delta_{i,j}\delta_{\ell,\ell'}\frac{1}{\Delta x}\delta(t-t')\, .
\end{equation}
Although these arguments are standard, the validity of the derivation 
is restricted.  Formally speaking, we focus on the limiting
case $L \to \infty$, $ d \to 0$, $\beta \to 0$, $d/\beta^2 
\simeq \alpha \simeq O(1)$  for the lattice model 
(\ref{time_evolution_operator_diffusion}). 
It is stimulating to analyze 
(\ref{time_evolution_operator_diffusion}) 
together with a renormalization group idea 
so as to derive a hydrodynamic description for general cases.

%%%%%%%%%%%%%%%%%%%%%%%%%%%%%%%%%%%%%
% acknowledgment                    %
%%%%%%%%%%%%%%%%%%%%%%%%%%%%%%%%%%%%%

\ack

This work was supported in part by grant-in-aid for scientific 
research Nos. 20115009 (JO), 21740283 (JO), 19540394 (SS), 
21015005 (SS) from the Ministry of Education, 
Culture, Sports, Science and Technology, Japan.

\appendix

\section{Path probability of particle number} \label{path-particle}

The probability of a particle-number trajectory 
$\{n(t)\}$ is also expressed 
within the Doi-Peliti formalism. We briefly explain the expression 
in this Appendix because there are no references mentioning 
the result explicitly. (See \cite{Andreanov2006} for 
a related discussion.) We start with an expression for the 
probability $P_n(t)$:
\begin{equation}
\fl\quad
P_{\nf} (\tau)=\lim_{\Delta t \to 0}
\left.
\left( 
\prod_{t=\Delta t}^{\tau-\Delta t} \sum_{n_t=0}^\infty 
\right)
\left(
\prod_{t=0}^{\tau-\Delta t} K(n_t \to n_{t +\Delta t})  
\right)
P_{\nint} (t)
\right\vert_{n_0=\nint,n_\tau=\nf}\, ,
\label{path-formal}
\end{equation}
where $K(n\to m)$ is equivalent to the previously-defined 
transition probability (\ref{transition_prob}) applied for 
a small time step $\Delta t$:
\begin{equation}
K(n \to m) =\frac{1}{m!}
\bra m \left\vert \rme^{-\hat H \Delta t} \right\vert n \ket\, ,
\end{equation}
and we used the most elementary decomposition of unity 
$\bm{1}= \sum_{n=0}^\infty | n \rangle \langle n |/n!$.
Rewriting $K(n \to m)$ as 
\begin{eqnarray}
K(n \to m) 
&=&  \sum_{\sigma=-\infty}^\infty
\frac{1}{(n+\sigma)!}
\bra n+\sigma  
\left\vert \e^{-\hat H \Delta t} 
\right\vert n \ket \delta_{m, n+\sigma} \nonumber \\
&=&  \int_{-\pi}^\pi \frac{\rmd \theta }{2\pi} 
\e^{- \rmi \theta(m-n)}
\sum_{\sigma=-\infty}^\infty  
\frac{1}{(n+\sigma)!}\, 
\e^{\rmi \theta \sigma}
\bra n+\sigma  \left\vert \e^{-\hat H \Delta t} \right\vert n \ket\, , 
\nonumber
\end{eqnarray}
we express the transition probability during a small
time interval $\Delta t$
as 
\begin{eqnarray}
K(n \to m) 
=  \int_{-\pi}^\pi \frac{\rmd \theta}{2\pi}
\exp 
\left[-\rmi \theta (m-n) 
-\sum_{\sigma} L(\rmi \theta, n;\sigma) \Delta t\right]\, ,
\label{K-exp}
\end{eqnarray}
where  $L(\rmi\theta,n;\sigma)$ was defined as 
\begin{equation}
L(\rmi\theta, n;\sigma)
=\frac{1}{(n+\sigma)!}
\bra n+\sigma  \left\vert \hat H \right\vert n \ket 
\rme^{\rmi \theta \sigma}\, . 
\label{L-exp}
\end{equation}
The expression (\ref{path-formal}) with (\ref{K-exp}) and (\ref{L-exp})
provides the probability of a trajectory $\{ n(t) \}$.  On the basis of 
this expression, one may develop a system-size expansion as we did in
section \ref{sec:new}. This formulation 
might be regarded as a sophisticated version
of the discussion in Ref. \cite{KMK}.

\end{document}